\documentclass[graybox]{svmult}

\usepackage{helvet}         
\usepackage{courier}        
\usepackage{type1cm}        
\usepackage{makeidx}         
\usepackage{graphicx}        
\usepackage{subcaption}
\usepackage{multicol}        
\usepackage[bottom]{footmisc}

\newcommand{\alphaLN}{\ensuremath{\alpha^{{\rm L}}_{{\rm N}}}}
\newcommand{\betaLN}{\ensuremath{\beta^{{\rm L}}_{{\rm N}}}}
\newcommand{\alphaHN}{\ensuremath{\alpha^{{\rm H}}_{{\rm N}}}}
\newcommand{\betaHN}{\ensuremath{\beta^{{\rm H}}_{{\rm N}}}}
\begin{document}

\title*{Principal and complementary series representations at the late-time boundary of de Sitter}
\titlerunning{At the late-time boundary of de Sitter} 
\author{Gizem \c{S}eng\"or and Constantinos Skordis}
\institute{Gizem \c{S}eng\"or \at CEICO, Institute of Physics of the Czech Academy of Sciences, Na Slovance 1999/2, 182 21 Praha 8, Czechia
	\email{sengor@fzu.cz}
\and Constantinos Skordis \at CEICO, Institute of Physics of the Czech Academy of Sciences, Na Slovance 1999/2, 182 21 Praha 8, Czechia \email{skordis@fzu.cz}}
\maketitle

\abstract{We demonstrate how free massive scalar fields in the set up that usually appears in early universe inflationary studies, correspond to the principal series and complementary series representations of the group SO(d+1,1) by introducing late-time operators and computing their two-point functions.}

\section{Introduction}
\label{sec:1}

Representations of Lie groups play an important role in quantum field theory since Wigner pointed out that elementary particles correspond to unitary irreducible representations of the isometry group of Minkowski spacetime, the Poincare group \cite{Wigner}. Of relevance to cosmology, is the de Sitter spacetime, a maximally symmetric spacetime with a positive cosmological constant, whose isometry group is also the conformal group of Euclidean space in one less dimension. The presence of conformal symmetries in de Sitter suggest the possibility to approach de Sitter physics in the framework of holography. Holography has been valuable tool for studies on  Anti de Sitter, a maximally symmetric spacetime with a negative cosmological constant and its applicability to the other maximally symmetric spacetimes is an active area of investigation. Following \cite{unitarity} and \cite{twopoint}, we summarize how to recognize the unitary irreducible representations of the de Sitter isometry group $SO(d+1,1)$ at the late-time boundary so as to gather clues on the inner workings of quantum field theory and the holographic nature of de Sitter spacetime.  

\section{The de Sitter geometry, the de Sitter group and the late-time boundary}
\label{sec:2}
The aim of this section is to introduce notation and some background information so as to establish connection between mathematics and physics literatures with focus on quantum field theory.

The de Sitter geometry is the vacuum solution to Einstein equations with a positive cosmological constant \cite{desitter}. In global coordinates, with the metric convention $diag(-,+,+,\dots)$, the metric for $d+1$ dimensional de Sitter is 
\begin{equation}
ds^2_{global}=g^{\mu\nu}_{global}dX^\mu dX^\nu=-dT^2+\frac{1}{H^2}cosh^2(HT)d\Omega^2_d,
\end{equation}
where $d\Omega^2_d$ denotes the metric on $d$ dimensional sphere and $H$ is the Hubble constant associated with the de Sitter scale ($l_{dS}=H^{-1}_{dS}$). The time coordinate runs in the range $T\in (-\infty, \infty)$. Due to the behaviour of the cosine hyperbolic function, this spacetime undergoes accelerated expansion in the interval $T\in [0,\infty)$. Within the cosmic history of our universe we have evidence of two epochs of accelerated expansion. One of these is the primordial epoch of inflation and the second is the current epoch of dark energy domination. Inflationary and dark energy epochs correspond to de Sitter like epochs and a lot of information about these epochs can be obtained by studying quantized fields on de Sitter. 

\begin{figure}
	\centering
	\begin{subfigure}{0.3\textwidth}
		\includegraphics[width=\textwidth]{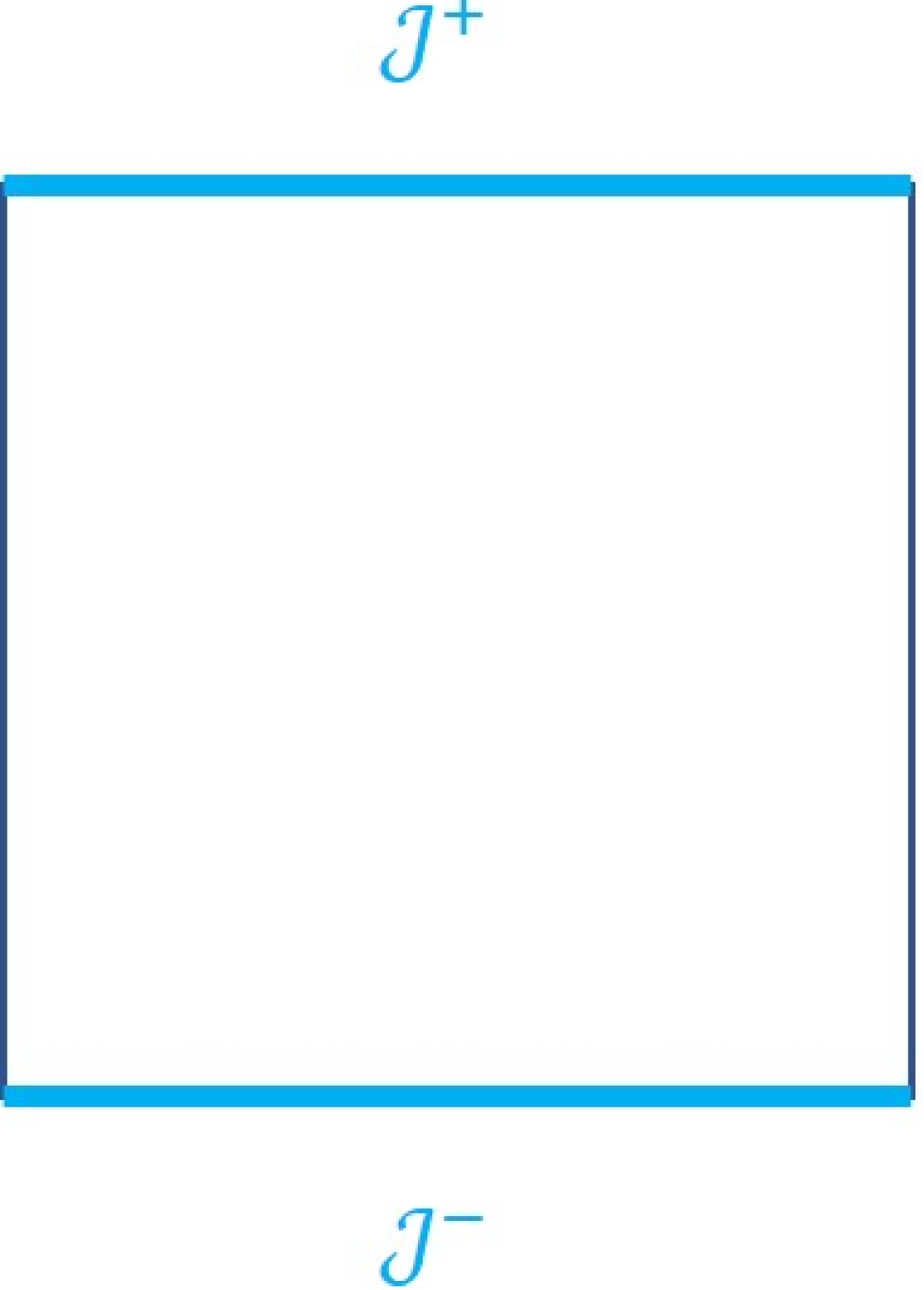}
		\caption{in global coordinates.}
		\label{fig:globalcoords}
		\end{subfigure}
	\begin{subfigure}{0.3\textwidth}
	\includegraphics[width=\textwidth]{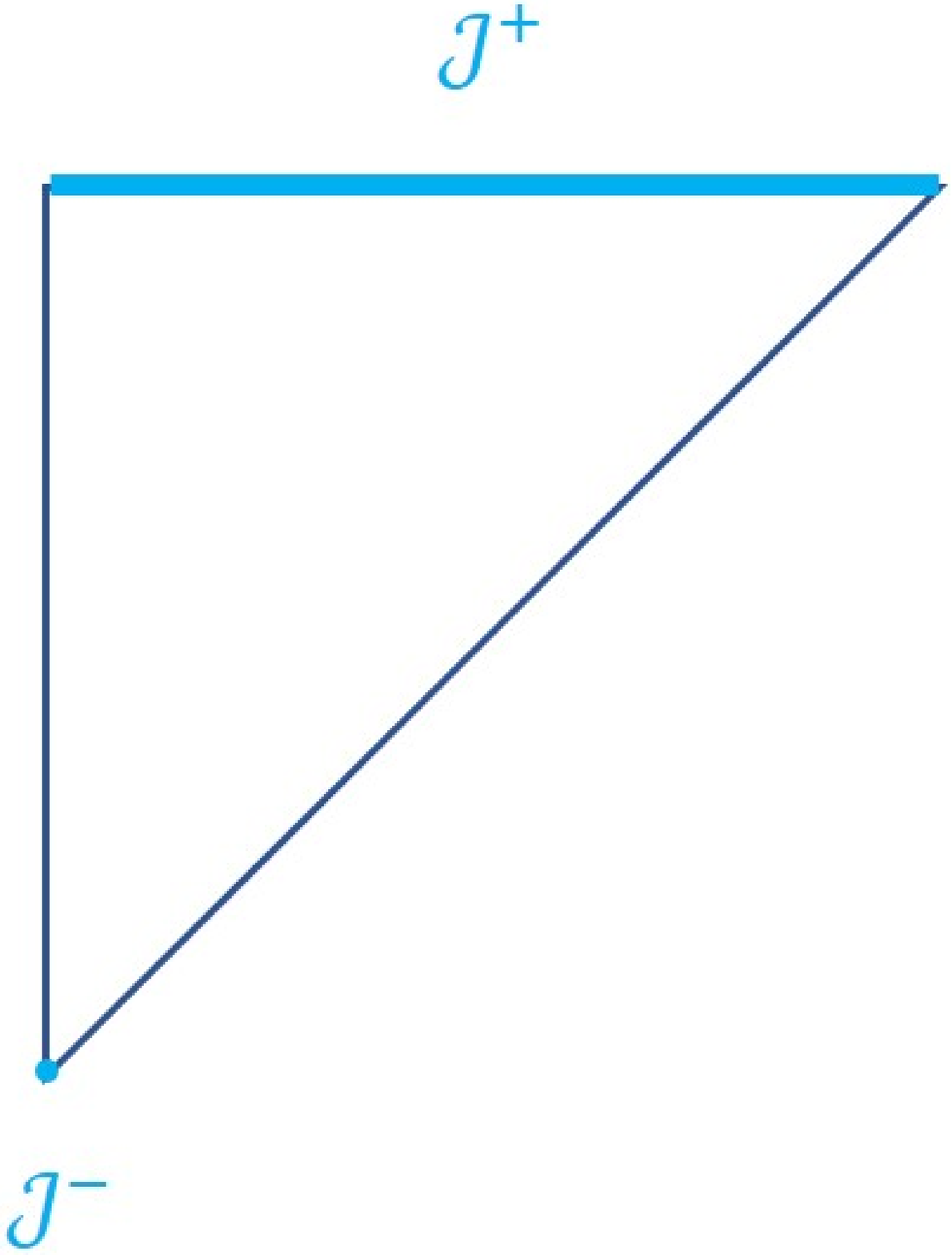}
	\caption{in planar coordinates.}
	\label{fig:planarcoords}
\end{subfigure}
\caption{Conformal diagrams for de Sitter}
\end{figure}

The de Sitter geometry has two conformal boundaries which lie along the time direction. These are referred to as the \emph{early-time boundary}, denoted by $\mathcal{I}^-$,  and the \emph{late-time boundary}, $\mathcal{I}^{+}$. Figure \ref{fig:globalcoords} shows these boundaries in light blue, for the conformal diagram of de Sitter in global coordinates. We will carry out our calculations in the so called planar patch or Poincare patch coordinates where the metric is
\begin{equation}
ds^2_{planar}=-dt^2+e^{2Ht}d\vec{x}^2=\frac{-d\eta^2+d\vec{x}^2}{H^2|\eta|^2}.
\end{equation} 
As shown in figure \ref{fig:planarcoords} these coordinates have access to the entire late-time boundary that we are interested in while they give access to only a single point from the early-time boundary. We will work in terms of conformal time $\eta$ which runs in the range $\eta\in(-\infty,0]$, and the late-time limit corresponds to the limit $\eta\to0$. This coordinate system is the one used in inflationary studies, where late-time correlation functions \cite{maldacenadscft}, mainly two-point and three-point functions, can be used to put observational constraints on inflationary interactions with comparison to cosmic microwave background radiation observations \cite{planck}. 


The isometries of $d+1$ dimensional de Sitter geometry correspond to the group $SO(d+1,1)$, also referred to as the de Sitter group. The representation theory of $SO(d+1,1)$ is well established in the mathematics literature initiating from the works of Harish-Chandra. Here we will follow the monograph \cite{Dobrev}. A recent short review can also be found in \cite{Sun}. 

 The de Sitter group is composed of linear transformations on the real $d+2$ dimensional vector space that leave the following quadratic form invariant
\begin{equation}
\xi^2=\xi\gamma\xi=\xi^2_1+\dots+\xi^2_{d+1}-\xi^2_0
\end{equation} 
where $\gamma$ denotes the metric on flat $d+2$ dimensional spacetime with the nonzero components being $\gamma_{ii}=\dots=\gamma_{d+1d+1}=-\gamma_{00}=1$. The de Sitter transformations include dilatations with the corresponding subgroup denoted by $A=SO(1,1)$, spatial rotations $M=SO(d)$, spatial translations $\tilde{N}$ and special conformal transformations $N$. Moreover there is the maximally compact subgroup $K=SO(d+1)$. An important feature of de Sitter spacetime that sets it apart from the other two maximally symmetric spacetimes, Minkowski and Anti de Sitter, is that the de Sitter group does not involve time translations. 

The representations of the de Sitter group are induced by the parabolic subgroup $P=NAM$. Under a dilatation where $\vec{x}\to\lambda\vec{x}$ an operator among the representations of the de Sitter group transform as
\begin{equation}
\mathcal{O}(\lambda\vec{x})=\lambda^{-\Delta}\mathcal{O}(\vec{x}).
\end{equation}
The exponent $\Delta$ is called the scaling dimension and for the group $SO(d+1,1)$ it has the following form
\begin{equation}
\Delta=\frac{d}{2}+c
\end{equation}
where $c$ is called the \emph{scaling weight} and it determines which category a given representation belongs to.

The eigenvalues of the quadratic Casimir operator are $\mathcal{C}=l(l+d-2)+c^2-\frac{d^2}{4}$
and therefore the representations are labeled by spin $l$ the label for representations of $M=SO(d)$ and the scaling weight $c$, denoted in a compact way as $\chi=[l,c]$. The unitary representations of the de Sitter group fall under four different categories where for each category the range of $c$ and the  well defined inner product is different. Representations with purely imaginary scaling weight belong to the \emph{principal series} representations, these are irreducible and have a straightforward inner product. Three different categories span unitary representations with real scaling weight, the irreducible \emph{complementary series} and  \emph{discrete series} representations and the reducible \emph{exceptional series} representations. The well defined inner product involves intertwining operators for these categories. The range of $c$ and the accompanying intertwining operator in the inner product differs for each of these categories. We refer the reader to \cite{Dobrev} for the definitions of appropriate intertwining operators and to \cite{unitarity} for a summary and a practical use of the intertwining operator in the case of complementary series representations.

\section{The late-time operators}
\label{sec3:latetimeops}
We will introduce late-time operators that correspond to free quantized scalar fields on de Sitter following \cite{unitarity}. Our assumption is that the scalar field does not effect the geometry, the metric is fixed to be the de Sitter metric. A free, massive scalar field on de Sitter has the following action
\begin{equation} 
\label{action}
S=\int d^dx d\eta \sqrt{-g}\left[-\frac{1}{2}g^{\mu\nu}\partial_\mu\phi\partial_\nu\phi-\frac{1}{2}m^2\phi^2\right]
\end{equation}
where $g$ denotes the determinant of the metric.

For convenience we will consider the fourier modes of the field, $\phi_{k}(\eta)$, in momentum space $\vec{k}$. For a quantized field, the mode decomposition involves \emph{annihilation}, $a_{\vec{k}}$, and \emph{creation}, $a^\dagger_{\vec{k}}$, operators. These operators obey the following nontrivial commutation relation
\begin{equation}
\label{aadaggercommu}
\left[a_{\vec{k}},a^\dagger_{\vec{k}'}\right] =(2\pi)^d\delta^{(d)}(\vec{k}-\vec{k}').\end{equation}
The annihilation operator annihilates the vacuum state $|0\rangle$, while the creation operator with a given momentum acting on the vacuum state creates a state with the specified momentum
\begin{equation}
a_{\vec{k}}|0\rangle=0,~~~a^\dagger_{\vec{k}}|0\rangle=|\vec{k}\rangle.
\end{equation}
The states created in this way are normalized with respect to 
\begin{equation}
\langle \vec{k}|\vec{k}'\rangle=(2\pi)^d\delta^{(d)}(\vec{k}-\vec{k}').\end{equation}
With these operators the scalar field is decomposed into its fourier modes as follows
\begin{equation}
\phi(\vec{x},\eta)=\int \frac{d^dk}{(2\pi)^d}\left[\phi_k(\eta)a_{\vec{k}}+\phi^*_{k}(\eta)a^\dagger_{-\vec{k}}\right]e^{i\vec{k}\cdot\vec{x}},
\end{equation}
where $*$ denotes complex conjugation and $\phi(\vec{x},\eta)$ is a real field while the modes $\phi_k(\eta)$ are complex valued.

We demand Bunch-Davies initial conditions \cite{bunchdavies}, which require the field to behave as if it was on Minkowski at early times. The mode functions that satisfy the equations of motion with these initial conditions are given in terms of Hankel functions. The solutions split into two branches depending on how heavy the mass of the field is with respect to the Hubble scale as follows 
\begin{equation}
for~ m<\frac{d}{2}H:~\phi^L_k(\eta)=|\eta|^{\frac{d}{2}}H^{(1)}_\nu(k|\eta|),~where~\nu^2=\frac{d^2}{4}-\frac{m^2}{H^2}
\end{equation}
we call these as light fields, and
\begin{equation}
for~m>\frac{d}{2}H:~\phi^H_k(\eta)=|\eta|^{\frac{d}{2}}e^{-\rho\pi/2}H^{(1)}_{i\rho}(k|\eta|),~where~\rho^2=\frac{m^2}{H^2}-\frac{d^2}{4}
\end{equation}
we call these as heavy fields. Our notation is such that $\nu$ and $\rho$ are real and positive.

In the late-time limit, as $\eta\to 0$, the field goes to
\begin{equation}
\label{latetimeop_intro}
\lim_{\eta\to0}\phi(\vec{x},\eta)=\int \frac{d^dk}{(2\pi)^d}\left[|\eta|^{\frac{d}{2}-\mu_p}\alpha^p(\vec{k})+|\eta|^{\frac{d}{2}+\mu_p}\beta^p(\vec{k})\right]e^{i\vec{k}\cdot\vec{x}}
\end{equation}
where the label $p$ indicates light and heavy $p=L,H$ and $\mu_L=\nu$, $\mu_H=i\rho$. The operators $\alpha^p(\vec{k})$ and $\beta^p(\vec{k})$ are the \emph{late-time operators}. They are build out of annihilation and creation operators. Moreover the late-time operator $\alpha^p(\vec{k})$ has momentum dependence $k^{-\mu_p}$ while $\beta^p(\vec{k})$ has $k^{\mu_p}$. Looking at the scaling dimensions of these operators we recognize that their scaling dimensions match the format of scaling dimensions for the unitary irreducible representations of the de Sitter group with scaling weights $c=\pm\mu_p$ \cite{unitarity}, with plus sign for the $\beta$ operator, minus for the $\alpha$ operator. 

We can define states by acting on the vacuum state with the late-time operators, such as
\begin{equation}
\label{definingstates}
|\alpha^p(\vec{k})\rangle\equiv \mathcal{N}^p\alpha^p(\vec{k})|0\rangle
\end{equation}
where $\mathcal{N}^p$ is the normalization to be determined and the same argument follows for $\beta^p(\vec{k})$. Such states build from light late-time operators with $c=\pm\nu$ are normalizable with respect to the complementary series inner product in the range $0<m<\frac{d}{2}H$ while such states build from any of the heavy late-time operators with $c=\pm i\rho$ are normalizable with respect to the principal series inner product. From these observations we can identify our normalized heavy late-time operators
\begin{eqnarray}\label{normalizedprincp}
		\alphaHN(\vec{k})=&\sqrt{\rho\pi \sinh(\rho\pi)}\left[-i\frac{\Gamma(i\rho)}{\pi}e^{-\rho\pi}a_{\vec{k}}+\frac{1}{\sinh(\rho\pi)\Gamma(1-i\rho)}a^\dagger_{-\vec{k}}\right]\left(\frac{k}{2}\right)^{-i\rho}\\
		\betaHN(\vec{k})=&\sqrt{\rho\pi \sinh(\rho\pi)}\left[\frac{e^{\rho\pi}}{\sinh(\rho\pi)\Gamma(1+i\rho)}a_{\vec{k}}+i\frac{\Gamma(-i\rho)}{\pi}a^\dagger_{-\vec{k}}\right]\left(\frac{k}{2}\right)^{i\rho}
\end{eqnarray}
 as the principal series representations of the de Sitter group, and the normalized light late-time operators 
\begin{eqnarray}
	\label{normalizedcomp}
		\alphaLN(\vec{k})=-i2^{\nu/2}\left[a_{\vec{k}}-a^\dagger_{-\vec{k}}\right]k^{-\nu},\\
		\betaLN(\vec{k})=2^{-\nu/2}\left[\frac{1+i\cot(\pi\nu)}{1-i\cot(\pi\nu)}a_{\vec{k}}+a^\dagger_{-\vec{k}}\right]k^{\nu}
\end{eqnarray}
as the complementary series representations of the de Sitter group \cite{unitarity},\cite{twopoint}. The $\alpha^p_N$ and $\beta^p_N$ have nontrivial commutation relations inherited from the commutation relation of the annihilation and creation operators.

\section{The late-time two-point functions}
\label{sec:two-point functions}
   
At this point it is straight forward to calculate the two-point functions defined as $\langle\mathcal{O}_1(\vec{k})\mathcal{O}_2(\vec{k})\rangle\equiv\langle0|\mathcal{O}_1(\vec{k})\mathcal{O}_2(\vec{k})|0\rangle$. We obtain the following list \cite{twopoint}
\begin{eqnarray} 
\nonumber&\langle 
\alphaLN(\vec{k})\alphaLN(\vec{k}')\rangle
=2^\nu k^{-2\nu}(2\pi)^d\delta^{(d)}(\vec{k}+\vec{k}'),\\
&\langle \betaLN(\vec{k})\betaLN(\vec{k}')\rangle=\frac{ k^{2\nu}}{2^\nu}\frac{1+i\cot(\pi\nu)}{1-i\cot(\pi\nu)}(2\pi)^d\delta^{(d)}(\vec{k}+\vec{k}'),
\\
\nonumber&\langle \alphaLN(\vec{k})\betaLN(\vec{k}')\rangle=-i(2\pi)^d\delta^{(d)}(\vec{k}+\vec{k}'),~
\langle \betaLN(\vec{k})\alphaLN(\vec{k}')\rangle=i\frac{1+i\cot(\pi\nu)}{1-i\cot(\pi\nu)}(2\pi)^d\delta^{(d)}(\vec{k}+\vec{k}'),
\end{eqnarray}
for the complementary series two-point functions, while for the principal series
\begin{eqnarray}
\nonumber&\langle\alphaHN(\vec{k})\alphaHN(\vec{k}')\rangle=-\frac{\Gamma(1+i\rho)}{\Gamma(1-i\rho)}e^{-\rho\pi}(2\pi)^d\delta^{(d)}(\vec{k}+\vec{k}')\left(\frac{k}{2}\right)^{-2i\rho},\\
&\langle\betaHN(\vec{k})\betaHN(\vec{k}')\rangle=i\rho\frac{\Gamma(-i\rho)}{\Gamma(1+i\rho)}e^{\rho\pi}(2\pi)^d\delta^{(d)}(\vec{k}+\vec{k}')\left(\frac{k}{2}\right)^{2i\rho}\\
\nonumber&\langle\alphaHN(\vec{k})\betaHN(\vec{k}')\rangle=e^{-\rho\pi}(2\pi)^d\delta^{(d)}(\vec{k}+\vec{k}'),~
\langle\betaHN(\vec{k})\alphaHN(\vec{k}')\rangle=e^{\rho\pi}(2\pi)^d\delta^{(d)}(\vec{k}+\vec{k}').
\end{eqnarray}
In \cite{twopoint} how these two-point functions contribute to field and conjugate momentum two-point functions at late-times which are related to observable quantities are discussed, in canonical and wavefunction quantization.

\bigskip

\begin{acknowledgement}
GŞ is supported by the European Union's Horizon 2020 research and innovation programme under the Marie Sk\l{}odowska-Curie grant agreement No 840709-SymAcc.
CS acknowledges support from the European Structural and Investment  Funds  and  the  Czech  Ministry  of  Education,  Youth  and  Sports  (MSMT)  
(Project  CoGraDS  -CZ.02.1.01/0.0/0.0/15003/0000437).
\end{acknowledgement}

\begin{thebibliography}{99.}%
%
%


%
%
%
\bibitem{bunchdavies}T. S. Bunch, P. C. W. Davies, Proc. Roy. Soc. Lond. A360 (1978) 117.
\bibitem{desitter}W. de Sitter, Mon. Not. Roy. Astron. Soc. 78 (1917) 3.
\bibitem{Dobrev} V. K. Dobrev, G. Mack, V. B. Petkova, S. G. Petrova and I. T. Todorov, \textit{Harmonic Analysis
on the n-Dimensional Lorentz Group and Its Application to Conformal Quantum Field
Theory}, Lect. Notes Phys. 63 (1977) 1.
\bibitem{maldacenadscft}J. M. Maldacena, JHEP 05 (2003) 013 [astro-ph/0210603]
\bibitem{planck} Planck collaboration, Astron. Astrophys.
641 (2020) A10 [1807.06211].
\bibitem{unitarity}S. Sengor, C. Skordis, JHEP 06 (2020) 041 arXiv:1912.09885 [hep-th].
\bibitem{twopoint}S. Sengor, C. Skordis, arXiv:2110.01635 [hep-th].
\bibitem{Sun} Z. Sun, arXiv:2111.04591 [hep-th].
\bibitem{Wigner}E. Wigner, Annals of
Mathematics 40 (1939) 149.
%
%
%
%
%
\end{thebibliography}
\end{document}